\documentstyle[12pt,psfig]{article}
\begin{document}
\title{Quenches and crunchs: 
Does the system explore in aging the same part 
of the configuration space explored in equilibrium ? }
\author{
Stefano Mossa$^{1,2}$, Giancarlo Ruocco$^{2}$,
\\ 
Francesco Sciortino$^{2}$, and Piero Tartaglia$^{2}$
\\
\it $^{1}$ Center for Polymer Studies and 
\\
\it Department of Physics,
\\
\it Boston University, Boston, Massachusetts 02215
\\
\it $^{2}$ Dipartimento di Fisica and 
\\
\it Istituto Nazionale
per la Fisica della Materia, 
\\
\it Universit\'a di Roma {\it La Sapienza},
\\
\it P.le Aldo Moro 2, I-00185 Roma, Italy.
}
\date{June 8, 2001}
\maketitle
\begin{abstract}
Numerical studies are providing novel information on the
physical processes associated to physical aging.
The process of aging has been shown to consist in a slow
process of explorations of deeper and deeper minima 
of the system potential energy surface. In this article we 
compare the properties of the basins explored in equilibrium 
with those explored during the aging process both for sudden temperature changes
and for sudden density changes. We find that 
the hypothesis that during the aging process
the system explores the part of the configuration space
explored in equilibrium holds only for shallow quenches or for the early aging dynamics. 
At longer times, systematic deviations are observed. In the case of 
crunches, such deviations are much more apparent.
\end{abstract}
\newpage
\section{Introduction}

At the glass transition temperature $T_g$ (Debenedetti 1997),
the characteristic relaxation 
time of a liquid becomes of the same order of the experimental time, thus
preventing equilibrium studies at lower $T$.
Material properties below $T_g$ depend on the
previous history (i.e. on the preparation technique, on
the cooling/compression rate and so on) as well as on 
the time spent in the glass-state. This 
time dependence, generically known as
physical aging, highlights the out-of-equilibrium (OOE)
condition of glasses and their extremely 
slow equilibration processes.
                   
Substantial amount of work has been
devoted to the understanding and to the formal description of the
supercooled liquid dynamics (G\"otze 1999, Cummins 1999), 
of the physics beyond the glass transition (Angell 1995) and of 
physical aging (Bouchaud {\it et al.} 1998, 
Cugliandolo {\it et al.} 1996, Latz 2000).
Numerical simulations of supercooled states, 
both in equilibrium and in controlled 
out-of-equilibrium conditions, have played an 
important role in the present developments (Kob 1999).
Although the time scales probed by numerical simulations 
are very different than experimental ones 
($100 ns$ compared to $s$), the numerical ``experiments'' 
appear to be able to reproduce features of real
materials (Utz {\it et al.} 2000, Barrat and Berthier 2001). 

The description of the aging dynamics as motion in configuration space,
has been very fruitful(Kob {\it et al.} 2000). It has been shown that, during aging,
the system explores deeper and deeper basins of the potential energy
surface.  The search for deeper basins during aging resembles the
exploration of deeper and deeper basins which takes place in
equilibrium on cooling. This similarity has been interpreted in term
of a decrease of the internal configurational temperature $T$ of the
system under aging (Kob {\it et al.} 2000).  
An analysis of the curvatures of the PES basins supported the
possibility that during the aging process the system visits the
same type of minima as the one visited in equilibrium.  This
analysis supported also the possibility of a 
thermodynamic description of the aging
system and a prediction of the internal configurational temperature in
full agreement with the numerical estimates (Sciortino and Tartaglia 2001a).

The outcome of these studies, which are still far from being settled,
suggest that the out-of-equilibrium glassy state,
notwithstanding its out of equilibrium condition, can be uniquely
determined by its kinetic temperature, its volume and the
properties of the basin in which the system is trapped. 
These quantities allow to develop a
thermodynamic description of the glass
state (Davies and Jones 1953, Nieuwenhuizen 1998, Speedy 1998, Mez\'ard
and Parisi 1999, Sciortino {\it et al.} 1999) which is currently tested in
experiments (Grigera and Israeloff 1999, Knaebel {\it et al.} 2000) 
and further simulations(Sciortino and Tartaglia 2001b).
.

From an experimental point of view, aging experiments are performed
at constant pressure. Moreover, often a change in pressure 
(and hence in density) is used to bring the system 
from an equilibrium to an out-of-equilibrium state. The change
in density produces a much more dramatic change in the PES than a
change in temperature. This opens the possibility that the
hypothesis of out-of-equilibrium dynamics as succession of
quasi equilibrium states may not apply to the crunch experiments
and hence that a more refined thermodynamic approach is requested for
glass materials produce with a crunch route. 

In this manuscript we revisit the 
hypothesis that during the aging process
the system explores the same part of the configuration space
by performing a more accurate analysis of the relation
basin-depth as a function of the  basin-curvature, made nowadays
possible by the increased computational facilities. 
We also present a comparison of this relation for different
quenching temperatures and for a crunch (Di Leonardo {\it et al} 2000).
\section{The models}   
We consider two models, an atomic one and a molecular one.

The first  microscopic model we consider is a binary
(80:20) mixture of Lennard-Jones particles (BMLJ), which in the following we
will call type $A$ and type $B$ particles.
The interaction between two particles of type $\alpha$ and $\beta$,
with $\alpha,\beta \in \{\rm A,B\}$, is given by $V_{\alpha\beta}=
4\epsilon_{\alpha\beta} [(\sigma_{\alpha\beta}/r)^{12}-
(\sigma_{\alpha\beta}/r)^6]$. The parameters $\epsilon_{\alpha\beta}$
and $\sigma_{\alpha\beta}$ are given by $\epsilon_{AA}=1.0$,
$\sigma_{AA}=1.0$, $\epsilon_{AB}=1.5$, $\sigma_{AB}=0.8$,
$\epsilon_{BB}=0.5$, and $\sigma_{BB}=0.88$.  The potential is
truncated and shifted at $r_{\rm cut}=2.5\sigma_{\alpha\beta}$.
$\sigma_{\rm AA}$ and $\epsilon_{\rm AA}$ are chosen as the unit of
length and energy, respectively (setting the Boltzmann constant $k_{\rm
B}=1.0$). Time is measured in units of $\sqrt{ m \sigma_{\rm AA}^2
/48\epsilon_{\rm AA}}$, where $m$ is the mass of the particles. 1000
particles were placed in a box of side 9.4. 
The studied isochore has density of 1.2.
To study the quenches,
we have equilibrated several independent configurations
at $T=T_i$ in the $NVT$ ensemble (Nos\'e-Hoover thermostat). Each of
the configuration has been quenched to $T_f$ by changing at $t=0$
the thermostat temperature to $T_f$. The thermostat constant is chosen
in such a way that, within 1000 MD steps, the average kinetic energy
thermalizes to $T_f$.
To study the crunches, we have equilibrated 280 independent 
configurations
at $T=0.35$ in the $NVT$ ensemble at density of 1.103. The coordinates
of each atom has been rescaled by 0.9714 at $t=0$ to
simulate a 10\% density increase. In this way, the final density of the
crunches coincides with the density of the quenches. 
Inherent structures of a system are calculated by conjugate gradient
algorithms. The minimization procedure is iterated until the
energy change is less than $10^{-15}$. The density of states is
then calculated by diagonalizing the corresponding Hessian matrix.

The second model we consider is a simple three-sites molecular model (LW), 
introduced by Wahnstr\"om and Lewis (Wahnstr\"om and Lewis 1993). 
The model is constructed by  gluing in a rigid molecule 
three identical Lennard Jones (LJ)
atoms. The shape of the molecule (an isoscele triangle) and the
LJ parameters were chosen to mimic as close as possible  
one of the most studied  glass-forming,  liquid,{\it ortho}-terphenyl.
The slow dynamics of this model has been recently revisited (A. Rinaldi
{\it et al.} 2001) in great details.
\begin{figure}[t]
\centering
\mbox{\psfig{figure=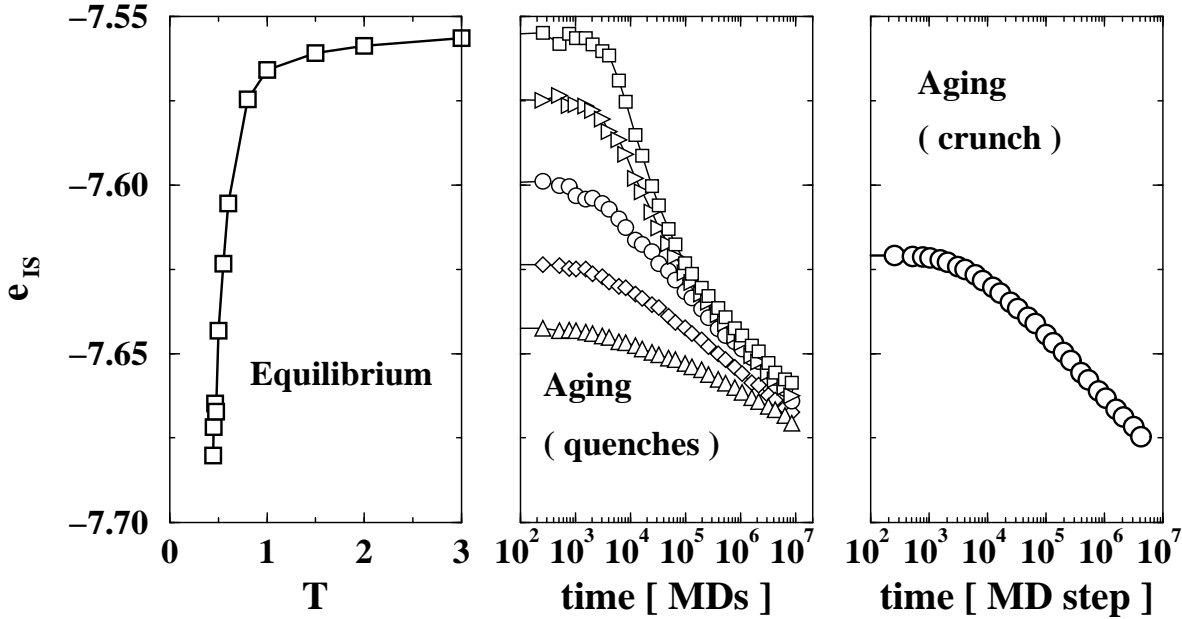,width=15cm,angle=0}}
\caption{Inherent structure energy as a function of $T$ in equilibrium
(left) and as a function of waiting time $t_w$ following a
quench (center) or a crunch (right).
In the central panel, $T_f=0.25$ and $T_i$ is $5$,$0.8$,$0.6$,$0.55$ and
$0.5$ from top to bottom. }
\label{fig:energy}
\end{figure}
For this model we present preliminary results for
a quench, starting from $T=480$ K down to $T=280$ K at constant density
$1108.4907$ Kg $/$ m$^3$. The simulated system is composed by 343 molecules. 
The integration time step is $0.01$ ps and
the aging dynamics is followed up to $1$ ns. Averages over more than 50
independent realizations are presented. 
\section{BMLJ model}
\subsection{Energies}
Following a quench or a crunch, the system finds itself in a
region of configuration space which is not explored in 
equilibrium under the externally imposed temperature and volume. 
The motion of the system in configuration space evolves in the
attempt of reaching the typical equilibrium configuration. 
A clear indicator of this evolution is the time evolution of the
potential energy. In the PES paradigm, the potential 
energy of the system can be expressed as
$E= e_{IS} + E_{vib}$, where $e_{IS}$ is the IS
energy and $E_{vib}$ describes the thermal excitations about the IS
(harmonic plus anharmonic vibrations). While the
vibrational energy does not show a significant aging dependence,
the time evolution of 
$e_{IS}$ provides a clear indication of the search for 
deeper and deeper potential energy minima, as shown in Fig.\ref{fig:energy}.  
\begin{figure}[t]
\centering
\mbox{\psfig{figure=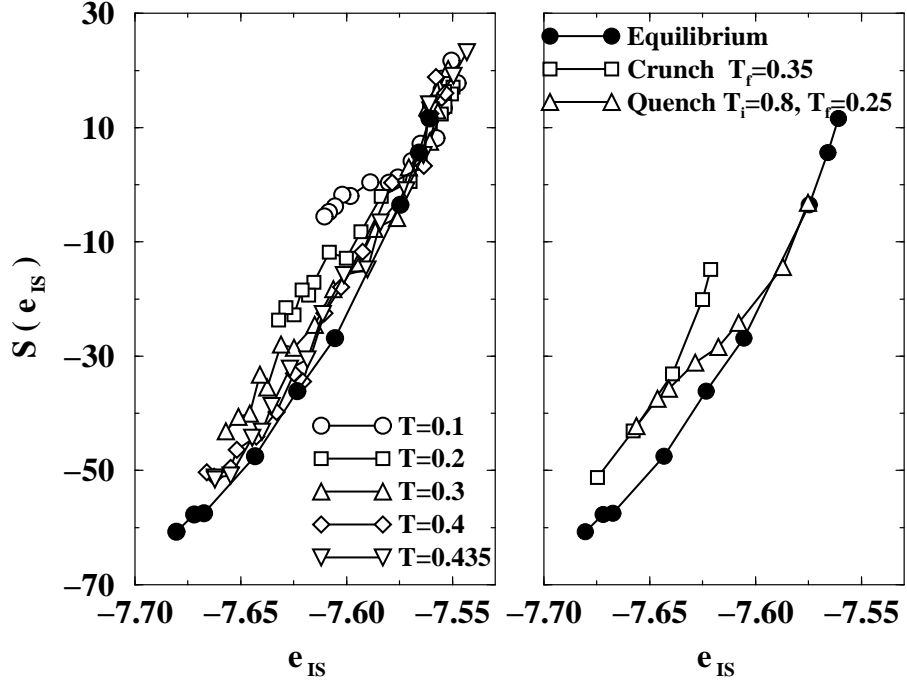,width=12cm,angle=0}}
\caption{Relation between $S \equiv \sum_{i=1}^{3N-3} log(\hbar
\omega_i/\epsilon_{AA})$   and $e_{IS}$. Data from [Kob {\it et al.}
2000] (left panel) and new data for crunch and quench aging (right panel).}
\label{fig:logwvseis}
\end{figure}
\subsection{Basins' curvatures} 
In equilibrium, below $T=1.$, the system start to explore deeper
and deeper basins, whose shape is depth dependent. 
A simple way to
characterize the shape of the PES basins  is to calculate the
density of states in harmonic approximation, expanding the
potential energy around the local minimum configuration.
The resulting distribution of frequencies characterizes, in
harmonic approximation, the volume in configuration space
associated to the basin. The density of states  allows
to estimate the vibrational free energy  $f_{basin}$ in harmonic approximation,
since
\begin{equation}
f_{basin}(T)=\frac{k_B T}{N} \sum_{i=1}^{3N-3} log( \beta \hbar \omega_i)
\end{equation}
A $T$-independent indicator of the basin shape can be defined as
$S \equiv \sum_{i=1}^{3N-3} log(\hbar \omega_i/\epsilon_{AA})$.

Studies on several models of liquids shows that along isochoric
paths, the depth dependence of the 
average curvature of the basins is model and density dependent.

In the case of the BMLJ at the studied density, the average shape of
the basins becomes wider and wider on moving to deeper and deeper basins,
as shown in Fig. \ref{fig:logwvseis}.  

Fig. \ref{fig:logwvseis} also shows $S(e_{IS})$ during the aging
dynamics following quenches and crunches, from different initial and
bath conditions. The important observation is that during 
$T$-changes, there is an initial part of the aging process were
the system explores in aging the same set of basins explored in
equilibrium. For longer times, deviations start to take place and the
system finds itself located in region of the configuration space
which are not commonly explored in equilibrium conditions. 
The effect is negligible (and indeed it was not noted  in previous studies)
at sufficiently high $T_f$ values (shallow quenches), but is becomes detectable
in deeper quenches. In future studies, it will become 
important to correlated the time at which the aging dynamics 
significantly separates from the equilibrium one with the
different mechanism of exploration of configuration spaces which characterize
also the dynamics of the system in equilibrium 
(i.e. saddle-dominated dynamics
 as compared to activated dynamics (Sciortino and Tartaglia 1997, La Nave {\it et al} 2000, Angelani {\it et al} 2000, Cavagna {\it et al} 2000)).
A related interesting question is the validity of the description of
the aging system as composed by two system in quasi-equilibrium at different 
temperatures (Sciortino and Tartaglia 2001) 
in conditions where the aging system explores 
basins which are not populated in equilibrium.

In the  crunch case, in the time window accessible to numerical simulations  
the system never explores region of configuration space
which are visited in equilibrium.
The starting configuration is 
different from any configuration visited in equilibrium. 
In this respect, the ensemble of configuration of the system after an initial
crunch can not be identified with an equilibrium ensemble 
at a different temperature, not even at infinite temperature.

Having said so, we call the reader attention on the fact that the
differences observed in $S$ are smaller than the entire 
variation of $S$ with aging. We also recall that the 
time windows
accessible to numerical experiment is such that the system, even after the
longest waiting time simulated, has not completely forgot the
initial configuration (or, in other words, the correlation functions
never decay to zero in the simulated time period). In this respect,
it is not that surprising that the basins explored in crunches are
different from the basins explored in aging and in equilibrium.
Finally we note that $S$, being the sum of logarithms, is driven by the
low frequency values. The small frequencies are the most affected by
size effects and by artificial localization of the eigenmodes.
Future studies should focus on the size effect on 
the data reported in Fig.\ref{fig:logwvseis}.

We note on passing that some model potentials, like for example the
BKS model for silica do not show, along an isochoric path any dependence
of the basin shape on the potential depth(Saika-Voivod {\it et al} 2001). 
This class of models
(and their corresponding materials) may provide simple 
aging and crunching dynamics compared to the BMLJ model here studied.
\begin{figure}[p]
\centering
\mbox{\psfig{figure=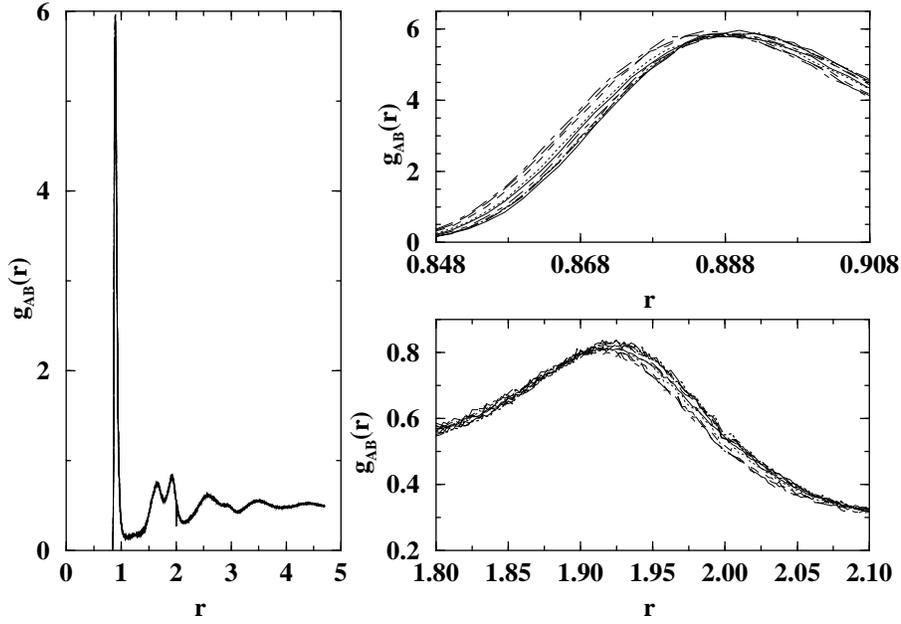,width=12cm,angle=0}}
\caption{
Radial distribution function for the AB atoms for eigth different temperatures
ranging from 2 to 0.446, evaluated in the IS configurations.
The right panels show the 
enlargement of two different $r$-regions. A significant enlargement is
requested to
highlight the very subtle changes accompanying the population of basins of
deeper and deeper IS energy.
}
\label{fig:gr}
\end{figure}
\begin{figure}[p]
\centering
\mbox{\psfig{figure=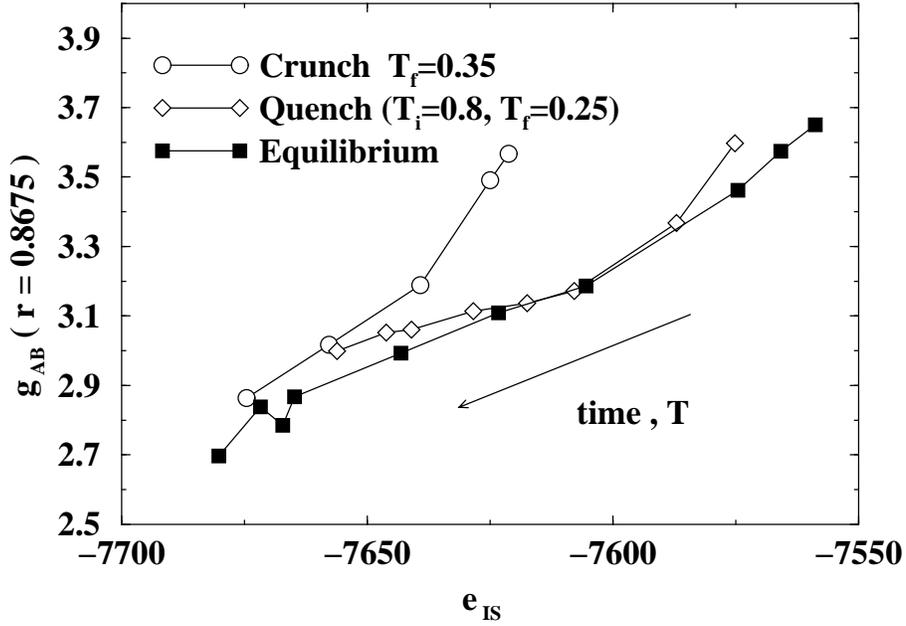,width=12cm,angle=0}}
\caption{
Dependence on $e_{IS}$ of $g_{AB}(r=0.8675)$ in equilibrium and
during aging (crunch and quench). Note that states with same
$e_{IS}$ value are realized with different pair distribution functions.
}
\label{fig:grovs}
\end{figure}
\subsection{Structure of the liquid in the inherent structures}
\begin{figure}[t]
\centering
\mbox{\psfig{figure=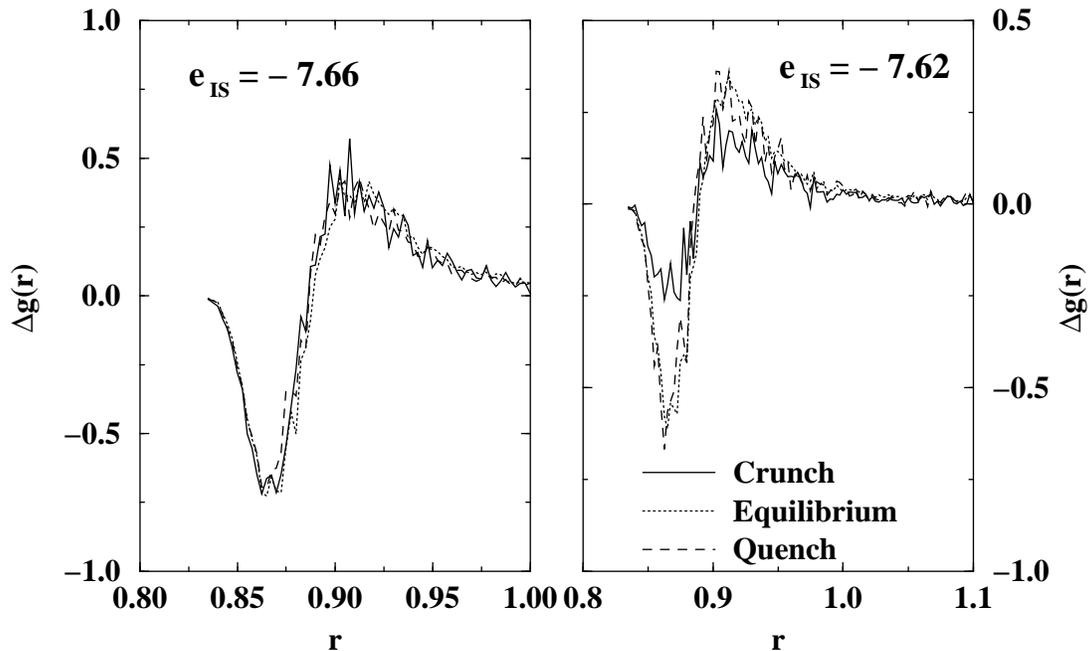,width=14cm,angle=0}}
\caption{Equilibrium, crunch and quench $g_{AB}$ at the same
$e_{IS}$ values. To maximize the differences, the equilibrium 
$g_{AB}(r)$ at $T=2.0$ has been subtracted to all curves.}
\label{fig:grdiff}
\end{figure}
The inherent structure energy is, by definition, the integral of the
pair potential over $r$, weighted by the radial distribution function
$g(r)$. The changes in inherent structure energy on cooling does reflect
changes in the local structure. In this section we
look in detains these differences, with the aim of characterizing
in a microscopic way the differences in states with the same
inherent structure but different density of states.

Fig. \ref{fig:gr} shows the AB radial distribution function 
in equilibrium as evaluated in the IS configurations, i.e. once the
thermal distortion has been subtracted. The structure of the
liquid changes in a very minor way and can be visualized only 
with a very fine resolution. The net effect of cooling in $g_{AB}(r)$ 
appears as a shift of less than 0.001 in the average interatomic 
distance. 

To emphasize the $T$-changes we show in Fig. \ref{fig:grovs}
the relation between the basin potential energy and $g(r)$ for a
fixed $r$ value. The data confirm that under crunch, the
local liquid order is different from the equilibrium ones. 

Fig.\ref{fig:grdiff} shows the enlargements of $g(r)$ in
equilibrium and contrast them with the one during aging at the same
$e_{IS}$ value. 
Again, we note that the quench configuration is closer to
the equilibrium one as compared with the crunch one.
\begin{figure}[p]
\centering
\mbox{\psfig{figure=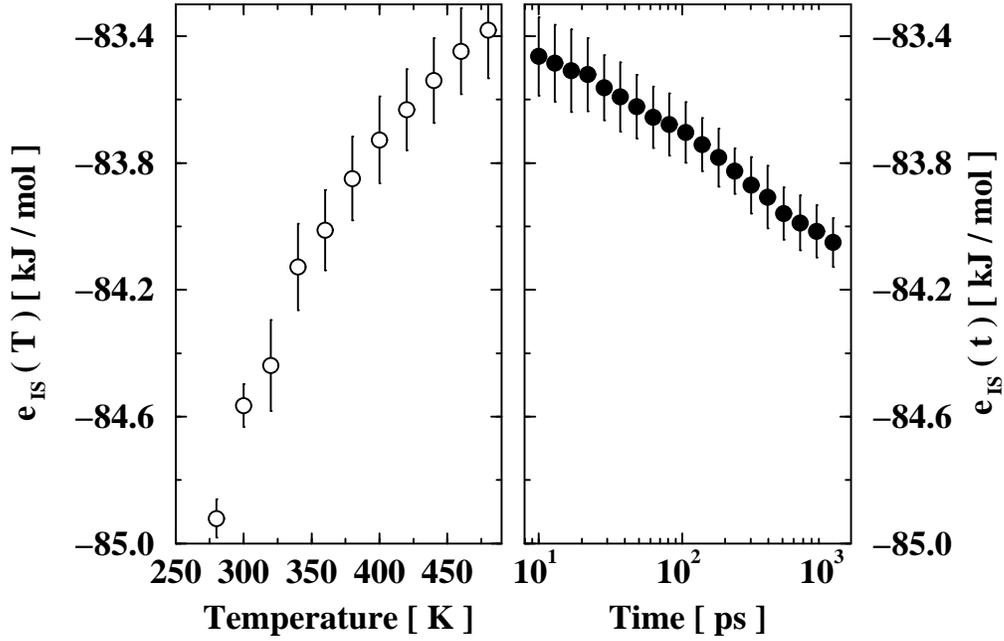,width=12cm,angle=0}}
\caption{Equilibrium $T$-dependence of the IS energy for the LW model
(left) and as a function of time during a quench from $T=480$ to $T=280 K$ (right).}
\label{fig:eisotp}
\end{figure}
\begin{figure}[p]
\centering
\mbox{\psfig{figure=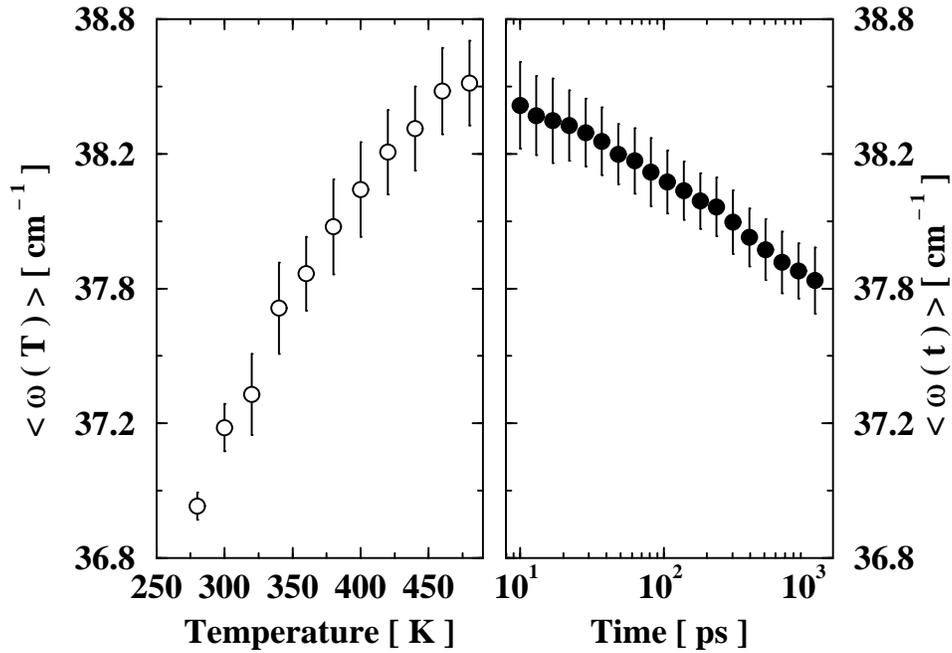,width=12cm,angle=0}}
\caption{Equilibrium $T$-dependence of the average frequency for the LW model
(left) and as a function of time during a quench from $T=480$ to $T=280 K$ (right)}
\label{fig:omotp}
\end{figure}
\section{LW model}
In the case of the LW model for ortotherphenyl
(Wahnstr\"om and Lewis 1993). , all results refer to
a quench case.
\subsection{Energies}
The $T$-dependence of the inherent structure energy for the
LW model is shown in Fig.\ref{fig:eisotp}, together with the
time evolution under aging. As for the BMLJ model, the aging dynamics
is characterized by a slow progressive reduction of the IS energy.
\subsection{Basins' curvatures} 
As in the BMLJ case, the basin curvature is correlated with 
the basin's depth. The $T$-dependence of the average
frequency in equilibrium and the $t_w$ dependence in aging is
shown in Fig.\ref{fig:omotp}.  
The frequency decreases on cooling or on aging as in the
BMLJ case. 

\begin{figure}[t]
\centering
\mbox{\psfig{figure=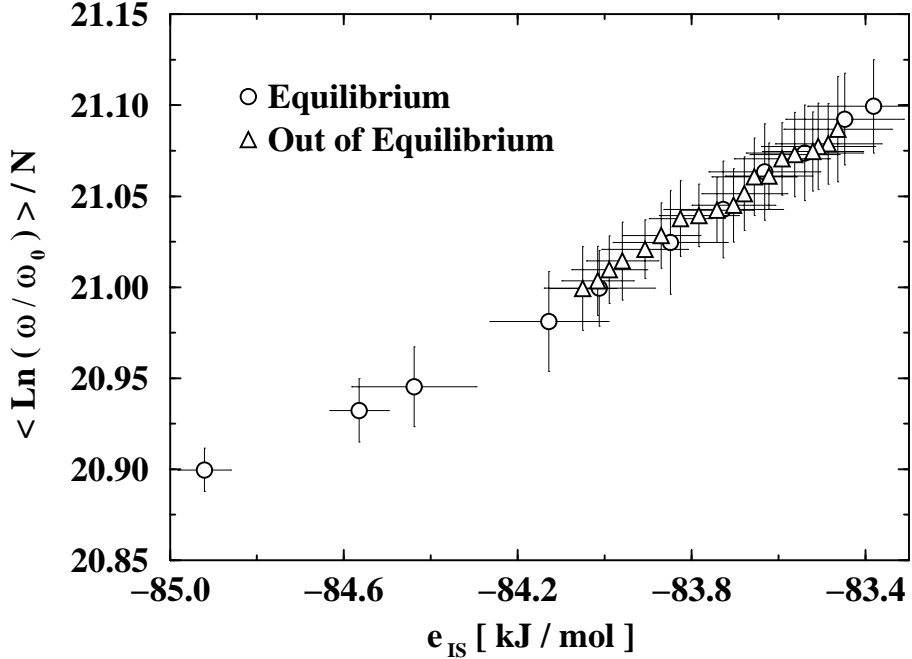,width=12cm,angle=0}}
\caption{Relation between $e_{IS}$ and the average basin curvature
for the LW potential, in equilibrium and in aging. ($\omega_0=1cm^{-1}$)}
\label{fig:isomotp}
\end{figure}

The $e_{IS}$ dependence of the local curvatures in equilibrium and
in aging is shown in Fig.\ref{fig:isomotp}. We note that for the
case of the LW potential, in the time window explored by the
numerical simulation and for the chosen $T_f$, 
the basins explored during aging coincides with
the basins explored in equilibrium.
\section{Conclusions}
In this manuscript we have compared the properties of
configuration space explored in equilibrium and
in out of equilibrium conditions, with the aim of 
deepening our understanding of the
physical mechanisms behind the aging process in disordered
materials.

A careful analysis of the relations between curvature
and depth of the potential energy basins reveals that
basins which are not statistically explored in equilibrium are
visited during the aging dynamics, especially during the
dynamics following a crunch. In quenches, such conditions appears to
hold also in the case of deep quench depth. At long times,
the hypothesis that during the aging process
the system explores the same part of the configuration space
does not seem to hold any longer, at least on the scale of
present day numerical calculations.

The differences in the basins appear to be located in the region of
very small frequencies and may not be clearly seen if
the average frequency is used as indicator of the basin curvature.
We have shown here that the average of the logarithm of the
frequency (a quantity which weigh more the very low
frequency spectrum) is indeed a better indicator. 
Such quantity is important since it quantifies the basin's
vibrational free energy. On the other hand, the very low
frequency part of the spectrum is very sensitive to size effects.
It could be that spurious localized modes
are stabilized by the limited size of the simulated system.
This call for a size-dependence detailed analysis 
of the relation between curvature and depth in equilibrium and
in aging.

Finally, we recall that the hypothesis that the
part of configurations space explored in aging and in equilibrium
are similar is an important element to be able to predict
the value of the internal temperature of the system and the 
associated response of the aging system to an external
perturbation. Indeed, in the case where such 
detailed comparison was performed, the basin shape and
curvature in aging and equilibrium do coincide.
Unfortunately, no evaluation of the internal temperature
has been performed yet for $t_w$ values where 
the possibility of predicting the value of $T_{int}$ should fail.
It would be very interesting to perform such accurate study
(i.e. the comparison between predictions and numerical calculations)
in the near future, in particular for crunches, where
the breaking of the assumption is apparent even at short waiting times.
\section*{Acknowledgements}
We are supported by INFM PRA-HOP, 
{\it INFM Parallel Computing Initiative} and MURST-COFIN2000.


\begin{thebibliography}{10} 
%
\bibitem{agingxxx} E.~Andrejew and J.~Baschnagel, Physica A {\bf 233},
117 (1996).
%
\bibitem{angelani}
L.~Angelani , G.~Ruocco, A.~Scala, and F.~Sciortino, 
Phys. Rev. Lett. {\bf 85}, 5356 (2000).
%
\bibitem{reviewgt} 
C.~A.~Angell, Science {\bf 267}, 1924 (1995). 
%
\bibitem{barrat} J.L. Barrat and L.Berthier, Phys. Rev. E 
{\bf 63}, 012503 (2001).
%
\bibitem{reviewaging} For a review see for example
J.~P. Bouchaud, L. F.  Cugliandolo, J. Kurchan, and M. M\'ezard,
in {\it Spin Glasses and Random Fields}, Edited by A.~P.~Young 
(World Scientific, Singapore, 1998), p.~161;
J. Kurchan, cond-mat/0011110, to be published in the special issue 
`Physics of Glasses' of the Comptes Rendus de Physique de l'Academie des Sciences.
%
\bibitem{cavagna}
K.~Bhattacharya , A.~Cavagna, A.~Zippelius, and I.~Giardina,
Phys. Rev. Lett. {\bf 85}, 5360 (2000).
%
\bibitem{letizia} 
L.~F.~Cugliandolo, J.~Kurchan, and P.~Le~Doussal, 
Phys. Rev. Lett. {\bf 76}, 2390  (1996).
%
\bibitem{cumminspisa} H. Cummins, 
{ J. Phys.: Condens. Matter} {\bf 11}, A95 (1999).
%
\bibitem{davies} R.O. Davies and G.O. Jones, { Adv.  in Physics} 
{\bf 2}, 370 (1953).
%
\bibitem{debenedetti} P.~G. Debenedetti, {\it Metastable Liquids}
(Princeton Univ. Press, Princeton, 1997).
%
\bibitem{dileonardo}
R.~Di Leonardo, L.~Angelani, G.~Parisi, and G.~Ruocco
Phys. Rev. Lett. {\bf 84}, 6054 (2000); L.~Angelani, 
R.~Di Leonardo, G.~Parisi, and G.~Ruocco, cond-mat/0011519, to be
published in Phys. Rev. Lett.
%
\bibitem{goetzepisa} W. G\"otze, { J. Phys.: Condens. Matter} {\bf 11}, A1 (1999).
%
\bibitem{grigera} T. S. Grigera and N. E. Israeloff, 
{Phys. Rev. Lett.} {\bf 83}, 5038 (1999).
%
\bibitem{expag}  
A.~Knaebel, M.~Bellour, J.~P.~Munch, V.~Viasnoff, 
F.~Lequeux  and J.~L.~Harden, Europhys. Lett., {\bf 52}, p. 73 (2000).
%
\bibitem{reviewsim} For a recent review on the contribution of
numerical simulation to the glass-transition phenomenon see 
for example: 
W.~Kob, J.  Phys.: Condensed Matter {\bf 11}, R85 (1999).
For out-of-equilibrium studies 
see also (Andrejew and Baschnagel 1996, Wahlen and Rieger 2000).
%
\bibitem{epl} W.~Kob, F.~Sciortino, and P.~Tartaglia,
{Europhys. Lett.} {\bf 49}, 590 (2000).
%
\bibitem{lanave}
E.~La~Nave, A.~Scala, F.~W.~Starr, F.~Sciortino, and H.~E.~Stanley,
Phys. Rev. Lett. {\bf 84}, 4605 (2000). 
E. La Nave, A. Scala, F.W. Starr, H.E. Stanley, F. Sciortino,
Phys. Rev. E {\bf 64},XXX (2001).
%
\bibitem{latz} A.~Latz, 
J. Phys.: Condensed Matter {\bf 12}, 6353 (2000);
A.~Latz, cond-mat/0106086.
%
\bibitem{parisi} M. M\'ezard and G. Parisi, 
{ J. Phys. Cond.  Matter} {\bf 11}, A157 (1999).
M. Cardenas, S. Franz and G. Parisi, {  J. Chem. Phys.} {\bf 110}, 1726 (1999).
%
\bibitem{teo} Th. M. Nieuwenhuizen,
{ Phys. Rev. Lett.} {\bf 80}, 5580 (1998); {\it ibid.} {\bf 79},
1317 (1997).
%
\bibitem{adele} A. Rinaldi, F. Sciortino and P. Tartaglia,
Phys. Rev. E {\bf 63}, 061210 (2001). 
%
\bibitem{st} F. Sciortino and P. Tartaglia,
Phys. Rev. Lett. {\bf 78}, 2385 (1997).
%
\bibitem{stk}  F. Sciortino, W. Kob and P. Tartaglia,
Phys. Rev. Lett.  {\bf 83} 3214 (1999).
%
\bibitem{st2} F. Sciortino and P. Tartaglia,
Phys. Rev. Lett.  {\bf 86}, 107 (2001). 
%
\bibitem{st3} F. Sciortino and P. Tartaglia,
J. Phys.: Condensed Matter {\bf XX}, XXX (2001).
%
\bibitem{robin} R. Speedy,
 { J. Phys.: Conds. Matter } {\bf 10}, 4185 (1998);  R. Speedy,
J. Phys. Chem. B {\bf 103}, 4060 (1999).
%
\bibitem{pabloaging} M. Utz, P. G. Debenedetti, 
F. H. Stillinger, Phys. Rev. Lett. {\bf 84}, 1471 (2000).
%
\bibitem{Agingsio2}  H.~Wahlen, H.~Rieger, J. Phys. Soc. Jpn. {\bf 69}, Suppl. A, 242 (2000).
%
\bibitem{lewis}
G.~Wahnstr\"{o}m and L.~J.~Lewis, Physica A {\bf 201}, 150 (1993);
L.~J.~Lewis and G.~Wahnstr\"{o}m, Solid State Comm. {\bf 86}, 295 (1993);
L.~J.~Lewis and G.~Wahnstr\"{o}m, J. of Non-Crystalline Solids {\bf 172-174},
69 (1994);
L.~J.~Lewis and G.~Wahnstr\"{o}m, Phys. Rev. E {\bf 50}, 3865 (1994),
G.~ Wahnstr\"om and L.~J.~Lewis, Prog. Theor. Phys. Suppl. {\bf 126},
261 (1997).
%
\end{thebibliography}
\end{document}